\documentclass[aps,prl,twocolumn,superscriptaddress,amsmath,amssymb,showpacs]{revtex4}
\usepackage{graphicx}
\usepackage{color}

\begin{document}

\title{Compression of ultrashort UV pulses in a self-defocusing gas}

\author{Luc Berg\'e}
\affiliation{CEA-DAM, DIF, F-91297 Arpajon, France}

\author{Christian K\"ohler}
\affiliation{Max-Planck-Institute for the Physics of Complex Systems, 01187 Dresden,
Germany}

\author{Stefan Skupin}
\affiliation{Max-Planck-Institute for the Physics of Complex Systems, 01187 Dresden,
Germany}
\affiliation{Friedrich-Schiller-University, Institute of Condensed Matter Theory and Solid State
Optics, 07743 Jena, Germany}

\date{\today}

\begin{abstract}
Compression of UV femtosecond laser pulses focused into a gas cell filled with xenon is reported numerically.
With a
large negative Kerr index and normal dispersion, xenon promotes temporal modulational instability (MI) which can be
monitored to shorten $\sim$ 100 fs pulses to robust, singly-peaked waveforms exhibiting a fourfold compression factor.
Combining standard MI theory with a variational approach allows us to predict the beam parameters suitable for efficient
compression. At powers $\leq 30$ MW, nonlinear dispersion is shown to shift the pulse temporal profile to the rear zone.

\end{abstract}


\maketitle

Since the nineties, considerable progress have been reported in the field of optical pulse compression \cite{Alfano:SLS:06}. From the first
achievements of few-cycle pulses \cite{Nisoli:ol:22:522}, the omnipresent idea has been to control spectral
broadening, in order to reach shorter light structures. This goal can be completed by post-compressing pulses
resulting from the interplay between self-phase modulation (SPM) and group-velocity dispersion (GVD) in hollow
waveguides and soliton compression devices, e.g., fibers. Other techniques have been proposed, using negative phase shifts based on cascaded quadratic nonlinearities, which can preserve most of the pulse energy \cite{Liu:ol:24:1777,Ashihara:josab:19:2505,Bache:josab:24:2752}. Alternatively, few-cycle filaments of light can be created from the balance between Kerr self-focusing and plasma defocusing at higher peak powers \cite{Hauri:apb:79:673,Skupin:pre:74:056604,Zair:oe:15:5394}. Frequency conversion processes, that enlarge more the spectrum, have also been
exploited to generate ultrashort pulses at UV wavelengths. Recently, Fuji {\it et al.} \cite{Fuji:ol:32:2481} produced
12-fs pulses at 260 nm with $\sim 10$ $\mu$J energy through the four-wave mixing of 400 and 800 nm pulses in
filamentation regime \cite{Theberge:prl:97:023904,Trushin:ol:32:2432}. The same scheme was numerically optimized to
compress pulses below 2 fs in low-pressure argon cell \cite{Berge:ol:33:750}.

Apart from the effective self-defocusing nonlinearity promoted by cascaded quadratic interactions \cite{Liu:ol:24:1777}, few attention has been paid to materials having a negative Kerr response. These exist however, such as helium and xenon, for
laser wavelengths in the range of $\sim238-249$~nm. In  \cite{Lehmberg:oc:121:78}, KrF laser light was used to
measure the Kerr index $n_2$ of xenon, which can attain important negative values close to a two-photon
resonance at 249.6 nm. In  \cite{Junnarkar:oc:175:447}, numerical evidence was given to the principle of pulse
compression in (1+1) dimensional hollow fibers filled with xenon, where pulses at 243 nm could be drastically shortened
with minimum energy losses. In higher dimensional systems, nonlinear defocusing can offer a rich variety of
dynamical patterns, e.g., ring dark solitons and vortices \cite{Kivshar:pr:298:81}. For a medium supporting normal
GVD, modulational instability (MI) can moreover develop in time, providing further potentiality for pulse compression.

In this Letter, we investigate new propagation regimes implying a defocusing nonlinearity without any guiding device.
$100$~fs Gaussian pulses at 243~nm central wavelength can be compressed in a cell of xenon at ambient pressure to about
25~fs along a stage
of modulational instability. These compressed structures support propagation ranges up to 1~m.
Because nonlinearity is highly dispersive in the UV, we take into account the full frequency dependency
of the third-order susceptibility
$\chi^{(3)}(-\omega;\omega^{\prime\prime\prime},\omega^{\prime\prime},\omega^{\prime})$, where
$\omega=\omega^{\prime\prime\prime}+\omega^{\prime\prime}+\omega^{\prime}$. Results obtained from such
a rigorous
approach are still compatible with those inferred from a classical nonlinear Schr{\"o}dinger (NLS) model. Qualitative
behaviors can be predicted by combining a variational method
\cite{Berge:pop:3:824} with the classical MI theory for plane waves \cite{Bespalov:jetp:3:307,Couairon:pop:7:193}.

We assume a linearly-polarized electric field $\sim {\cal E}
\mbox{e}^{i k_0 z - i \omega_0 t} + c.c.$, where ${\cal E}(r,z,t)$
denotes its envelope and $I =
|{\cal E}|^2$ is the pulse intensity. The wave number $k_0 = n_0 \omega_0/c$ involves the linear refractive index of the
gas, $n_0 \simeq 1$, at central frequency $\omega_0$.  Because the maximum intensity $I_{\rm max}$ never exceeds 0.5
TW/cm$^2$ in the coming simulations, the peak density of free electrons created by ionization always remains below
$10^{14}$ cm$^{-3}$, rendering plasma generation negligible. The physics can thus be described by the following envelope equation in Fourier space
\begin{equation}
\label{chi3model}
\begin{split}
\partial_z\hat{\cal E} & = \left[ \frac{i}{2k(\omega)r} \partial_r r \partial_r
+i\left(k(\omega)-k_0-k^{(1)}\tilde{\omega} \right)\right]\hat{\cal E}\\
& \quad + \frac{3i\omega^2}{2c^2k(\omega)}\iint 
\chi^{(3)}(-\omega;\omega-\omega^{\prime\prime}-\omega^{\prime},\omega^{\prime\prime},\omega^{\prime})\\
& \quad \times \hat{\cal E}^*(\tilde{\omega}^{\prime}+\tilde{\omega}^{\prime\prime}-\tilde{\omega})
\hat{\cal E}(\tilde{\omega}^{\prime\prime}) \hat{\cal E}(\tilde{\omega}^{\prime})
d\omega^{\prime\prime}d\omega^{\prime},
\end{split}
\end{equation}
where $r = \sqrt{x^2 + y^2}$ and $\tilde{\omega}=\omega-\omega_0$ denotes the envelope frequency. Here, $z$ is the propagation variable and $t$ a
retarded time.
Linear dispersion for xenon is included via $k(\omega)$ \cite{Dalgarno:procrsoca:259:424}, while the expression for the
nonlinear susceptibility $\chi^{(3)}$ is that given in \cite{Junnarkar:oc:175:447} [see also Fig.~\ref{fig4}(b)].
For sufficiently narrow spectral bandwidths, nonlinear as well as high-order linear dispersion, self-steepening and
space-time focusing operators are expected to have a minor influence on the qualitative pulse dynamics.
We will thus compare results obtained from Eq.~(\ref{chi3model}) with those from the
classical NLS
equation
\begin{equation}
\label{NLS}
\partial_z {\cal E} = \frac{i}{2k_0} r^{-1} \partial_r r \partial_r {\cal E}  - i\frac{k^{(2)}}{2} \partial_t^2 {\cal E}
 + i\frac{\omega_0}{c} n_2 |{\cal E}|^2 {\cal E},
\end{equation}
with $k^{(2)}=13.12$~fs$^2$/cm \cite{Dalgarno:procrsoca:259:424} and 
$n_2 = - 1.56 \times 10^{-17}$~cm$^2$/W \cite{Junnarkar:oc:175:447}. Both propagation models are integrated numerically
for Gaussian pulses with input power $P_{\rm in}$, beam waist $w_0$ and 1/e$^2$ pulse
half-width $\tau_p$. The delocalizing action of the nonlinearity is balanced by linearly focusing the incident beam. For
notational convenience, we will employ the definition of the critical power for the self-focusing of Gaussian beams,
$P_{\rm cr} \simeq \lambda_0^2/2\pi n_0 |n_2| \simeq 6$ MW. Key processes should thus be transverse diffraction, normal GVD and Kerr
defocusing. This allows us to capture the nonlinear dynamics by means of two different, simple analytical techniques.

On the one hand, we perform a two-scale variational analysis \cite{Berge:pop:3:824} resulting into the dynamical system
\begin{equation}
\label{variational}
\frac{k_0^2 w^3 w_{zz}}{4} = 1 + \frac{p \tau_p}{\sqrt{2} \tau}; \,\,\,\frac{\tau^3 \tau_{zz}}{4 k^{(2)}} = k^{(2)} - \frac{p \tau_p \tau}{\sqrt{2} k_0 w^2},
\end{equation}
where $p = P_{\rm in}/P_{\rm cr}$. This system governs the beam waist [$w(z)$] and 1/e$^2$ temporal radius [$\tau(z)$]
of Gaussian pulses, starting from $w(0) = w_0$ and $\tau(0) = \tau_p$. The normalized on-axis intensity behaves as
$I/I_0 = w_0^2 \tau_p/[w^{2}(z) \tau(z)]$ and the lens action can be modelled through suitable phase contributions
containing $d_z w(0) = - w_0/f$. Such an approximation method cannot describe the fine spatio-temporal deformations of
the pulse. Nevertheless, it usually yields estimates of the maximum intensity and of the pulse scales in space and time,
which support the comparison with direct numerical results. As seen from Eq.~(\ref{variational}), the pulse dynamics
differs from
standard self-focusing by the sign in front of the nonlinear term $\sim p$. When the pulse duration does not vary too
much, e.~g., for weak nonlinearities, the equation for $w(z)$ can readily be integrated and yields a focus reached at
the distance $z_{\rm min} \simeq f/[1 + (f/z_0)^2(1 + p/\sqrt{2})$], $z_0 \equiv \pi n_0 w_0^2/\lambda_0$ being the
Rayleigh length of the input beam. At high enough powers, the pulse can, in contrast, undergo significant compression in
time, as the Kerr response competes with normal GVD. This property will be exploited below.

On the other hand, the standard stability analysis for plane waves can bring further insight into the pulse dynamics. We
assume that the highest intensity zones of the pulse serve as plane-wave distributions, as long as local perturbations
have typical wavelengths (periods) less than the size (duration) of the background field with intensity close to its
maximum, $I_{\rm max}$. We can thus linearize Eq.~(\ref{NLS}) for perturbations
oscillating with the transverse wavenumber $k_{\perp}$ and frequency $\overline{\omega}$
\cite{Bespalov:jetp:3:307}, so that the MI growth rate expresses as $\gamma = \mbox{Re} \,( \Omega  \sqrt{2 \omega_0 
|n_2| I_{\rm max}/c - \Omega^2})$, where $\Omega^2 = k^{(2)}\overline{\omega}^2/2 - k_{\perp}^2/2 k_0$. Modulational
instability develops for positive values of $\Omega^2$ only. Following \cite{Couairon:pop:7:193}, a necessary condition
for MI is that the optimum perturbation wavenumber $k_{\perp}^{\rm max}$ and frequency $\overline{\omega}_{\rm max}$, linked to
each other by
\begin{equation}
\label{3}
\overline{\omega}_{\rm max} \simeq \sqrt{\frac{2 \omega_0  |n_2|  I_{\rm max}}{k^{(2)} c} + \frac{(k_{\perp}^{\rm max})^2}{k_0 k^{(2)}}}, 
\end{equation}
must satisfy $k_{\perp}^{\rm max} > \sqrt{2} \pi/w_{0}$ and $\overline{\omega}_{\rm max} > \sqrt{2} \pi/\tau_{p}$ at given $I_{\rm
max}$. MI fully takes place whenever perturbation frequencies $\overline{\omega} \geq \overline{\omega}_{\rm max}$ exist. Fixing
the value of $k_{\perp}^{\rm max} \sim \sqrt{2} \pi/w_{\rm min}$, where $w_{\rm min}$ is the smallest beam waist reached
at focus, it is then sufficient to evaluate whether the frequency range $\sqrt{2} \pi/\tau_p \leq \overline{\omega} \leq
\overline{\omega}_{\rm up} = \sqrt{2} \pi/\tau_{\rm min}$ fulfills the condition $\overline{\omega}_{\rm up} > \overline{\omega}_{\rm max}$, to conclude
if MI is efficiently seeded. Here, $\overline{\omega}_{\rm up}$ includes the minimum pulse duration $\tau_{\rm min}$. We conjecture
that reliable estimates for $I_{\rm max}$, $w_{\rm min}$ and $\tau_{\rm min}$ are provided by the variational equation
(\ref{variational}). In the following, these quantities will be used to evaluate $k_{\perp}^{\rm max}$,
$\overline{\omega}_{\rm up}$ and
$\overline{\omega}_{\rm max}$.

\begin{figure}
\includegraphics[width=8.5cm]{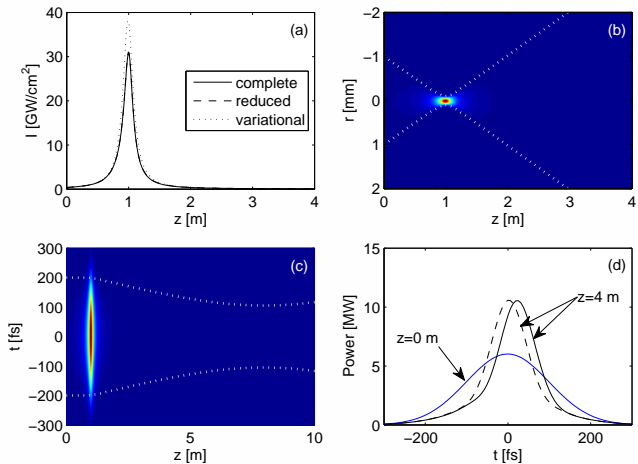}
\caption{(a) Peak intensities of a Gaussian pulse with $P_{\rm in}=6$~MW, $w_0=1$~mm and $\tau_p=200$~fs, focused in
xenon at atmospheric pressure ($f=1$~m). Solid line corresponds to the complete model Eq.~(\ref{chi3model}), dashed
(superimposed) line to Eq.~(\ref{NLS}), and dotted line to the variational model Eq.~({\ref{variational}}). (b)
Spatial and
(c) temporal dynamics of the full model. White dots reproduce the beam waist and 1/e$^2$ temporal extent computed from
Eq.~({\ref{variational}}). (d)  Power profiles after 4~m of propagation for the full model (solid line) and the reduced
model Eq.~(\ref{NLS}) (dashed line). The blue line shows the initial pulse.}
\label{fig1}
\end{figure}

An example for a stable configuration is shown in Fig.~\ref{fig1}, where a Gaussian pulse with $P_{\rm in} = 6$~GW,
$w_0=1$~mm
and $\tau_p = 200$ fs is simulated. From our variational model we find $I_{\rm max} \simeq 40$~GW/cm$^2$, $w_{\rm
min} \simeq 100~\mu$m and $\tau_{\rm min}\simeq 100$~fs, resulting in $k_{\perp}^{\rm max}\simeq 444$~cm$^{-1}$,
$\overline{\omega}_{\rm up} \simeq 0.044$~fs$^{-1}$ and $\overline{\omega}_{\rm max} \simeq 0.29$~fs$^{-1}$.
Since $\overline{\omega}_{\rm up}<\overline{\omega}_{\rm max}$, we expect stability of the temporal stripe designed by
the pulse in the $(t,z)$ plane.
The peak intensity
[Fig. \ref{fig1}(a)], computed from direct simulations (solid and dashed curves) and from Eq.~(\ref{variational})
(dotted curve), reaches a
maximum near the focal distance $z = 1$ m. Here, the intensity curves accounting or not for nonlinear dispersion are
almost identical. Figure \ref{fig1}(b) shows the spatial profile computed from Eq. (\ref{chi3model}). The beam focuses
then
diffracts with a divergence angle given by $\tan \theta_{\rm beam} \approx \lambda_0/\pi w_{\rm min}$. Figure
\ref{fig1}(c) illustrates the evolution of the on-axis temporal profile in the $(t,z)$ plane, which confirms robustness of the temporal profile against MI.
The white dots reproduce the functions $w(z),\tau(z)$ computed from Eq.~(\ref{variational}). In Fig.~\ref{fig1}(d), the minimal pulse
duration of about 100~fs is reached after a propagation distance of 4~m for both Eqs.~(\ref{chi3model}) and (\ref{NLS}),
while the same extent in time is attained from the variational model at $z=8$~m.
 
\begin{figure}
\includegraphics[width=8.5cm]{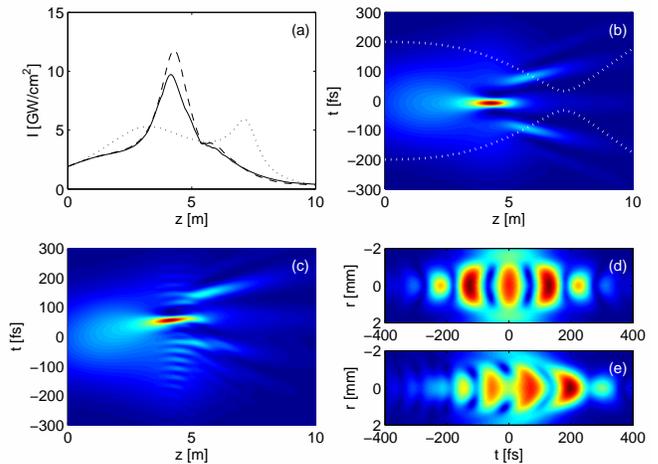}
\caption{(a) Peak intensities of the same Gaussian pulse as in Fig.~\ref{fig1}, but with $P_{\rm in}=30$~MW and $f=5$~m, using the same plotstyle as in Fig.~\ref{fig1}(a). (b) Temporal on-axis dynamics obtained from Eq.~(\ref{NLS}); white dots
correspond to the variational approximation. (c) Same from the complete model Eq.~(\ref{chi3model}). Spatio-temporal intensity distribution at
$z=8$~m showing MI from (d) the reduced model, (e) the complete model.}
\label{fig2}
\end{figure}

Let us then inspect regions of temporal instability. When we increase both the input power and focal length by a factor
five, the variational model predicts $I_{\rm max} \simeq 6$~GW/cm$^2$, $w_{\rm
min} \simeq 640~\mu$m and $\tau_{\rm min}\simeq 33$~fs, resulting in $k_{\perp}^{\rm max}\simeq 69$~cm$^{-1}$,
$\overline{\omega}_{\rm up} \simeq 0.13$~fs$^{-1}$ and $\overline{\omega}_{\rm max} \simeq 0.071$~fs$^{-1}$.
Since $\overline{\omega}_{\rm up} > \overline{\omega}_{\rm max}$, nonlinearities should thus seed MI and compression in time.
This is actually confirmed by the simulations shown in Fig.~\ref{fig2}. Compression leads to a short peak of $\sim
25$~fs duration at $z = 4$~m, before MI
fully breaks up the pulse temporal distribution.
The modulation period is about $2
\pi/\overline{\omega}_{\rm max} \sim 150$ fs, which agrees with Figs.~\ref{fig2}(d) and (e).
Note that again the minimum duration
predicted by Eq.~(\ref{variational}) occurs
beyond the compression stage seen in the simulations. The numerical pulse is subject to leaks of power, which
the variational approach cannot describe by preserving the input power within a Gaussian ansatz \cite{Berge:pop:3:824}. However,~Eq. (\ref{variational})
reproduces
compression rates agreeing well with the numerical data. While nonlinear dispersion clearly influences the pulse dynamics by moving its temporal centroid towards positive times,
we observe qualitative agreement between results obtained from Eqs.~(\ref{chi3model}) and (\ref{NLS}) [compare
Figs.~\ref{fig2}(b), (c) and \ref{fig2}(d), (e)].

\begin{figure}
\includegraphics[width=8.5cm]{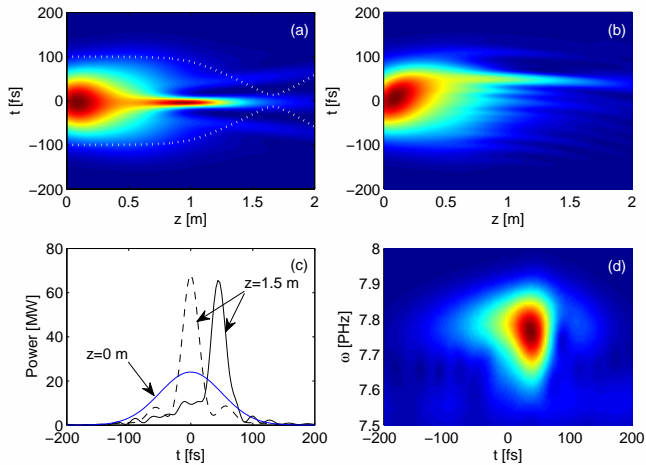}
\caption{Dynamics of a Gaussian pulse with $P_{\rm in}=24$~MW, $w_0=0.3$~mm and $\tau_p=100$~fs, focused in xenon with $f=1$~m. (a) Temporal on-axis dynamics obtained from the reduced model; white dots correspond
to variational results. (b) Same from the full model. (c) Power profiles after 1.5~m of propagation for the full model
(solid line) and the reduced model (dashed line). The blue line shows the initial pulse. (d) XFROG trace at $z=1.5$~m
using a $25$~fs reference pulse.} 
\label{fig3}
\end{figure}

\begin{figure}
\includegraphics[width=8.5cm]{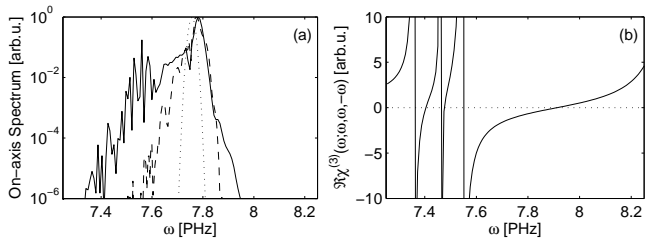}
\caption{(a) On-axis normalized spectra for the pulses shown in Fig.~{\ref{fig2}}(e) (dashed
line) and Fig.~{\ref{fig3}}(c) (solid line). The dotted line represents the spectrum of the 100 fs pulse at $z = 0$.
(b) Real part of the normalized nonlinear susceptibility $\chi^{(3)}$ in the degenerated case (see Ref.
\cite{Junnarkar:oc:175:447}).}
\label{fig4}
\end{figure}

Figure \ref{fig3} concerns the main result of this Letter. By tuning the input parameters
appropriately, MI can be
handled, such that only one modulation mainly affects the pulse temporal profile. Since a negative $n_2$ may lead to
compression at distances $z > z_{\rm min}$, the pulse can then not only reach very short durations, but also keep them
over long distances. The condition for this regime of optimal pulse compression is obviously $\overline{\omega}_{\rm up}
\simeq \overline{\omega}_{\rm max}$.
Let us examine the evolution of a Gaussian pulse, for which we reduce both initial
waist, pulse duration and power to $w_0 = 0.3$~mm, $\tau_p = 100$~fs, and $P_{\rm in}=24$~MW, respectively, to obtain
$\overline{\omega}_{\rm max} \simeq 0.14$~fs$^{-1}$ and $\overline{\omega}_{\rm up} \simeq
0.17$~fs$^{-1}$.
Figures \ref{fig3}(a) and (b) detail the pulse
evolution computed from Eqs.~(\ref{NLS}) and (\ref{chi3model}), respectively. MI is weakly seeded and the
100~fs pulse undergoes three modulations, among which only the central one is amplified and compressed along
further propagation down to $\sim 25$~fs. The resulting structure appears to behave like a 1D
sech-soliton in time, capable of preserving high intensity and short durations over at least 50~cm. Again we observe a slight shift to the trailing part of the pulse caused by the nonlinear dispersion. At such weak intensities, we can also infer that two-photon absorption remains of small influence. It is nevertheless important to point out that, in contrast to filamentary compression in self-focusing regime
\cite{Hauri:apb:79:673,Skupin:pre:74:056604,Zair:oe:15:5394}, short pulses obtained with the above compression scheme
are homogeneous in radial direction. Hence, compression applies not only to the on-axis
intensity profiles, but also to the pulse power profiles. Moreover, the XFROG trace shown in Fig.~\ref{fig3}(d) indicates that the compressed pulses are
nearly transform limited.

A last point concerns spectral broadening, as SPM is responsible for frequency variations $\Delta \omega \simeq 
k_0 \Delta z |n_2|  I_{\rm max}/\tau_{\rm min}$ around the focus. Spectral variations should remain confined
within the narrow bandwidth of 238--249~nm (7.57--7.92~PHz), in order to preserve a negative value of 
$n_2\sim\chi^{(3)}(\omega;-\omega,\omega,\omega)$ along the optical path
[see Fig.~\ref{fig4}(b)]. Figure \ref{fig4}(a) confirms that the spectrum has a small
bandwidth of $\Delta \omega
< 0.1$~PHz at half-maximum. For the simulation shown in Fig.~\ref{fig3}, spectral wings reach the range
where $n_2$ is not necessarily negative, but the dominant peak always remains located around the central frequency
[solid line in
Fig.~\ref{fig4}(a)].
We can notice that the usual broad low-intensity supercontinuum ($\Delta\omega_{\rm SC}/\omega_0 > 1$) routinely
observed from the propagation of plasma-induced filaments and caused by the mechanism of intensity clamping is missing.
We indeed find $\Delta \omega_{\rm SC}/\omega_0 < 0.1$, where $\Delta\omega_{\rm SC}$ is measured at $10^{-5}$ times
the maximum spectral intensity. Supercontinuum generation is thus
inhibited, as the peak intensity stays at moderate values.

Further increase of the input power to 60 MW leads to the occurrence of strong MI and multiple
pulse splitting ($\overline{\omega}_{\rm up} > \overline{\omega}_{\rm max}$, not shown here). For such high
powers, however, a major
part of the broadened spectrum falls into the range of multi-photon resonances below 7.57~PHz, and strong discrepancies
between the solutions of Eq.~(\ref{chi3model}) and Eq.~(\ref{NLS}) appear.
Consequently, while it is possible in the NLS model to produce even shorter pulses $<20$~fs by simply increasing the
initial peak power, dispersion of the nonlinear
susceptibility prevented us from achieving shorter durations with Eq.~(\ref{chi3model}). Such propagation regimes will
be addressed in a future publication.

In summary, numerical simulations have highlighted the dynamics of UV pulses focused into a self-defocusing gas cell.
Ultrashort optical structures can naturally be formed in (3+1)-dimensional media with negative $n_2$ and normal dispersion, and propagate over long ranges without experiencing a wide spectral broadening. By mixing simple analytical procedures, we proved that this new compression mechanism
follows from modulational instability in time, which can be controlled to optimize the self-compression process over
long distances. To end with, we find it worth emphasizing that nonlinear dispersion yielding variations in the Kerr
index of at least one order of magnitude does not significantly change results compared to propagation models
assuming a constant susceptibility. Here, we indeed showed that at moderate powers
deviations associated with nonlinear dispersion remain limited and their qualitative effect is similar to 
classical pulse self-steepening in focusing media.

This work has been performed using HPC resources from GENCI-CCRT/CINES (Grant 2009-x2009106003).

\end{document}